\documentclass[utf8]{frontiersSCNS} 
\usepackage{url,hyperref,lineno,microtype,subcaption}
\usepackage[onehalfspacing]{setspace}
\usepackage{subcaption}
\usepackage{booktabs}
\usepackage{commath}
\usepackage{multirow}

\usepackage{journal_macros} 

\def\keyFont{\fontsize{8}{11}\helveticabold}
\def\firstAuthorLast{J. Ruohotie {et~al.}} 
\def\Authors{J. Ruohotie\,$^{1}$*, E.~K.~J.~Kilpua\,$^{1}$, S.~W.~Good\,$^{1}$, M.~Ala-Lahti\,$^{1,2}$}
\def\Address{$^{1}$Department of Physics, University of Helsinki, Helsinki, Finland \\
$^{2}$Department of Climate and Space Sciences and Engineering, University of Michigan, Ann Arbor, MI, USA\\

}
\def\corrAuthor{Julia Ruohotie}

\def\corrEmail{julia.ruohotie@helsinki.fi}

\begin{document}
\onecolumn
\firstpage{1}

\title[Small flux ropes in sheaths]{\textbf{Small-scale flux ropes in ICME sheaths}} 

\author[\firstAuthorLast ]{\Authors} 
\address{} 
\correspondance{} 

\extraAuth{}

\maketitle


\begin{abstract}

Sheath regions of interplanetary coronal mass ejections (ICMEs) are formed when the upstream solar wind is deflected and compressed due to the propagation and expansion of the ICME. Small-scale flux ropes found in the solar wind can thus be swept into ICME-driven sheath regions. They may also be generated locally within the sheaths through a range of processes. This work applies wavelet analysis to obtain the normalized reduced magnetic helicity, normalized cross helicity, and normalized residual energy, and uses them to identify small-scale flux ropes and Alfv\'{e}n waves in 55 ICME-driven sheath regions observed by the Wind spacecraft in the near-Earth solar wind. Their occurrence is investigated separately for three different frequency ranges between $10^{-2} - 10^{-4}$ Hz. We find that small scale flux ropes are more common in ICME sheaths than in the upstream wind, implying that they are at least to some extent actively generated in the sheath and not just compressed from the upstream wind. Alfv\'{e}n waves occur more evenly in the upstream wind and in the sheath. This study also reveals that while the highest frequency (smallest scale) flux ropes occur relatively evenly across the sheath, the lower frequency (largest scale) flux ropes peak near the ICME leading edge. This suggests that they could have different physical origins, and that processes near the ICME leading edge are important for generating the larger scale population. 

\tiny\keyFont{ \section{Keywords:} Magnetic fields, Interplanetary coronal mass ejections, Solar wind, Flux ropes, Near-Earth space} 
\end{abstract}


\section{Introduction}
\label{sec:intro} 

Small-scale flux ropes (SFRs) are heliospheric structures commonly observed in the solar wind at variety of heliospheric distances \citep[e.g.,][]{moldwin2000,Feng2007,Feng2008,Kilpua2009,Cartwright2010,Yu2014,Yu2016,Hu2018,Chen2020SFR,Murphy2020,Chen2021SFR,Chen2022SFR,Zhao2020,zhao2021}. Similar to the large-scale flux ropes observed in a subset of interplanetary coronal mass ejections \citep[ICMEs; e.g.,][]{Kilpua2017,good2020b}, SFRs feature organized changes in the magnetic field direction, but with shorter durations and scale sizes. Previous studies typically define SFRs to have durations of a few tens of minutes to a few hours, and scale sizes of a few hundred Earth radii. Similar to large-scale ICME flux ropes, SFRs tend to have enhanced magnetic fields and low plasma beta with respect to their surroundings \citep[e.g.,][]{Yu2014}, but they usually lack signatures of expansion, i.e. their speed profiles are relatively flat, the magnetic field magnitude does not systematically peak near the middle of the structure, and the proton temperature is typically not less than the expected solar wind temperature for a given solar wind speed \citep[e.g.,][]{Yu2014}. As a consequence, there are also a significant population of SFRs with plasma beta $\sim 1$. There is now evidence that the occurrence rate of SFRs is at least slightly higher closer to the Sun at the Earth's orbit \citep{Cartwright2010,zhao2021}.

The origin of SFRs is not clear, and it is likely that SFRs have multiple sources. One suggestion is that some SFRs originate from small solar eruptions, i.e. weak flares and narrow CMEs \citep[e.g.,][]{Rouillard2011}. This is supported by the finding that narrow CMEs exhibit flux rope structure in multi-viewpoint white-light coronagraph observations \citep[e.g.,][]{Sheeley2009}, and that the duration and size of SFRs have power law scalings \citep[e.g.,][]{Feng2007,Hu2018}, similar to solar flares and CMEs \citep[e.g.,][]{lu1991,Pant2021}. However, their power indices differ, so the connection is not straightforward. Different indices could hint at differences in the underlying physical processes releasing them \citep[e.g.,][]{Hu2018}. Another possible source for SFRs at the Sun is the quasi-periodic release of plasma from the tips of coronal streamers. Heliospheric imager observations have allowed tracing of these streamer blobs (when they are entrained at the leading edges of high speed streams) to interplanetary space, where they are identified as a repeating pattern of density enhancements and flux ropes \citep{SanchezDiaz2017}. Interplanetary origins have also been invoked. For example, SFRs could form by reconnection near the heliospheric current sheet (HCS) \citep[e.g.,][]{Cartwright2010} or be self-generated via MHD-scale turbulence  \citep[e.g.,][]{Zheng2018,Hu2018}. 

SFRs have been detected in all types of solar wind, but the studies cited above show that they are most abundant in the slow solar wind. Recent studies have also highlighted that, when swept into an ICME-driven sheath region, SFRs can contribute to the energization of charged particles \citep{Kilpua2021energetic}. When impacting the Earth, they have also been shown to trigger magnetospheric substorms \citep{kim2017}. Since ICME sheaths are primarily plasma and field gathered from the slow solar wind, it is expected that SFRs are frequently present in ICME sheaths. It is also possible that SFRs are actively generated within sheaths. 

In this paper, we report the findings of a statistical study of SFRs identified in 55 ICME sheath regions detected in the near-Earth solar wind by the Wind spacecraft. SFRs have been identified by their normalized cross-helicity, residual energy and magnetic helicity signatures, similar to the analysis conducted by \cite{Zhao2020} using Parker Solar Probe data. The occurrence and magnetic field fluctuation properties of SFRs in ICME sheaths and in the preceding solar wind are investigated and compared. 

The paper is organized as follows: In Section~\ref{sec:datamethods}, data and methods are described, with results presented in Section~\ref{sec:results}. Findings are summarized and discussed in Section~\ref{sec:discussion}. 


\section{Spacecraft Data and Methods} 
\label{sec:datamethods}    

\subsection{Data} 
\label{sec:data}    

This work has used magnetic field data at 3~s time resolution from the Magnetic Fields Investigation \citep[MFI;][]{lepping1995} instrument onboard the Wind spacecraft \citep{ogilvie1997}.  Solar wind plasma parameters were obtained from Wind's Three-Dimensional Plasma and Energetic Particle Investigation \citep[3DP;][]{Lin1995}, also at 3~s resolution. The data were acquired through the NASA Goddard Space Flight Center Coordinated Data Analysis Web (CDAWeb) database.

\subsection{Methods} 
\label{subsec:methods}

In this work, we study 55 sheath regions driven by ICMEs that were observed between April 2004 and December 2015. The sheath events have been selected from the ICME catalogue by \citet{Nieves_Chinchilla2018} and HELCATS ICME catalogue. Obscure events and those with significant data gaps were removed, as well as events were the boundaries of the sheath were not clear. In a few cases the sheath boundaries have been moved to better match the shock or the ICME leading edge. The sheaths had a mean duration of 9 hrs 57 min, with the shortest lasting 50 min and the longest 26 hrs 27 min. Each sheath was preceded by a shock.

The SFR identification method follows that presented in \cite{Zhao2020}, and is based on methods first developed and applied by \cite{Telloni2012}. First, we calculated the normalized reduced magnetic helicity, $\sigma_m$, normalized cross-helicity, $\sigma_c$, and normalized residual energy, $\sigma_r$, using a wavelet transform \citep{Torrence1998} with the Paul wavelet function. Threshold criteria for identifying SFRs (described below) were defined using these three parameters.

Rotation in the magnetic field is measured with $\sigma_m$. The reduced form of $\sigma_m$ can be calculated from single spacecraft measurements \citep{Mattheus1982}. It may be written as
\begin{equation}
    \sigma_m = \frac{2 \operatorname{Im} \left[ W^{\ast}_y(\nu, t) \cdot W_z(\nu, t) \right] }{|W_x(\nu, t)|^2 + |W_y(\nu, t)|^2 + |W_z(\nu, t)|^2}
\end{equation}
where $\nu$ is the frequency associated with the wavelet function, and $W_x(\nu, t)$, $W_y(\nu, t)$, and $W_z(\nu, t)$ are the wavelet transforms of the magnetic field components \citep{Zhao2020}, here in GSE coordinates.

To distinguish SFRs from Alfv\'{e}n waves, $\sigma_m$ alone is not sufficient; $\sigma_c$ and $\sigma_r$ are also needed. $\sigma_c$ is a measure of the $\boldsymbol{v}-\boldsymbol{B}$ correlation (high correlation or anti-correlation for Alfv\'{e}n waves, weak correlation for SFRs) while $\sigma_r$ indicates whether more energy is found in magnetic or velocity fluctuations (equal energy in idealised Alfv\'{e}n waves, excess magnetic energy in SFRs). Cross-helicity and residual energy can be defined using the Elsässer variables, $\boldsymbol{z}^\pm = \boldsymbol{v} \pm \boldsymbol{b}$, where $\boldsymbol{v}$ is the velocity, $\boldsymbol{b} = \boldsymbol{B} / \sqrt{\mu_0 n_p m_p}$ is the magnetic field in velocity units, $n_p$ is the proton number density, and $m_p$ is the proton mass. Thus,
\begin{equation}
    \sigma_c = \frac{W^+(\nu, t) - W^-(\nu, t)}{W^+(\nu, t) + W^-(\nu, t)}
\end{equation}
and
\begin{equation}
    \sigma_r = \frac{2 \operatorname{Re} \left[ \mathcal{W}^{\ast}(z^+_x) \cdot \mathcal{W}(z^-_x) + \mathcal{W}^{\ast}(z^+_y) \cdot \mathcal{W}(z^-_y) + \mathcal{W}^{\ast}(z^+_z) \cdot \mathcal{W}(z^-_z) \right]}{W^+(\nu, t) + W^-(\nu, t)}
\end{equation}
where 
\begin{align}
    W^+(\nu, t) = |\mathcal{W}(z^+_x)|^2 + |\mathcal{W}(z^+_y)|^2 + |\mathcal{W}(z^+_z)|^2 \\
    W^-(\nu, t) = |\mathcal{W}(z^-_x)|^2 + |\mathcal{W}(z^-_y)|^2 + |\mathcal{W}(z^-_z)|^2
\end{align}
\citep{Zhao2020}.

SFRs typically display enhanced $|\sigma_m|$, low $|\sigma_c|$ and highly negative $\sigma_r$. We use the same identification criteria for SFRs as in \cite{Zhao2020}, which are as follows: 1) $|\sigma_m|>0.7$, 2) $|\sigma_c|< 0.4$, and 3) $\sigma_r < -0.5$. In this study, SFRs are identified in the frequency range $10^{-2} - 10^{-4}$~Hz (1.67--166.7 min) both in the sheaths and upstream solar wind. For solar wind at a speed of 500~km/s, this frequency range is equivalent to spatial scales of $5 \cdot 10^{4}$ -- $5 \cdot 10^{6}$ km (7.8 -- 780 $R_E$). Within this frequency range, three sub-ranges were examined separately: 1) high frequency (small period), $10^{-2}$ -- $2.16 \cdot 10^{-3}$~Hz (1.67 -- 7.71 min); 2) medium frequency (medium period),  $2.16 \cdot 10^{-3}$ -- $4.64 \cdot 10^{-4}$~Hz (7.71 -- 35.9 min); and low frequency (higher period), $4.64 \cdot 10^{-4}$ -- $10^{-4}$~Hz (35.9 -- 166.7 min). These sub-ranges have been chosen such that they have equal widths in logarithmic space.

Magnetic field fluctuation amplitudes and compressibility \citep[e.g.,][]{chen2015,good2020a,Kilpua2021stat} are also calculated within the identified SFRs, and compared to the non-SFR parts of the sheath and upstream wind. Fluctuation are defined as $\delta \mathbf{B} = \mathbf{B}(t) - \mathbf{B}(t+\tau)$, where $\tau$ is the time delay between two samples and $\delta B = |\delta \mathbf{B}|$ is the fluctuation amplitude. The normalized fluctuation amplitude is $\delta B/B$, where the field magnitude $B$ is calculated over the interval $[t,t+\tau]$. The compressibility of magnetic fluctuations is defined as $\delta |B| /\delta B$. 

Some structures that appear to be flux ropes from visual inspection of magnetic field data alone could in fact be Alfv\'{e}n waves \citep{Marubashi2010}. The wavelet analysis method has also been applied to identify Alfv\'{e}n waves, by changing the $\sigma_c$ and $\sigma_r$ criteria. Alfv\'{e}n waves usually have high $\sigma_c$ values and $\sigma_r$ close to zero. The following criteria for identifying Alfv\'{e}n waves have been used: 1) $|\sigma_m|>0.7$ (the same as for SFR identification), 2) $|\sigma_c| > 0.7$, and 3) $|\sigma_r| < 0.4$.

\section{Results}
\label{sec:results}      

\subsection{Example events}
\label{sec:examples}

\begin{figure}   
\centerline{\includegraphics[width=0.99\textwidth,clip=]{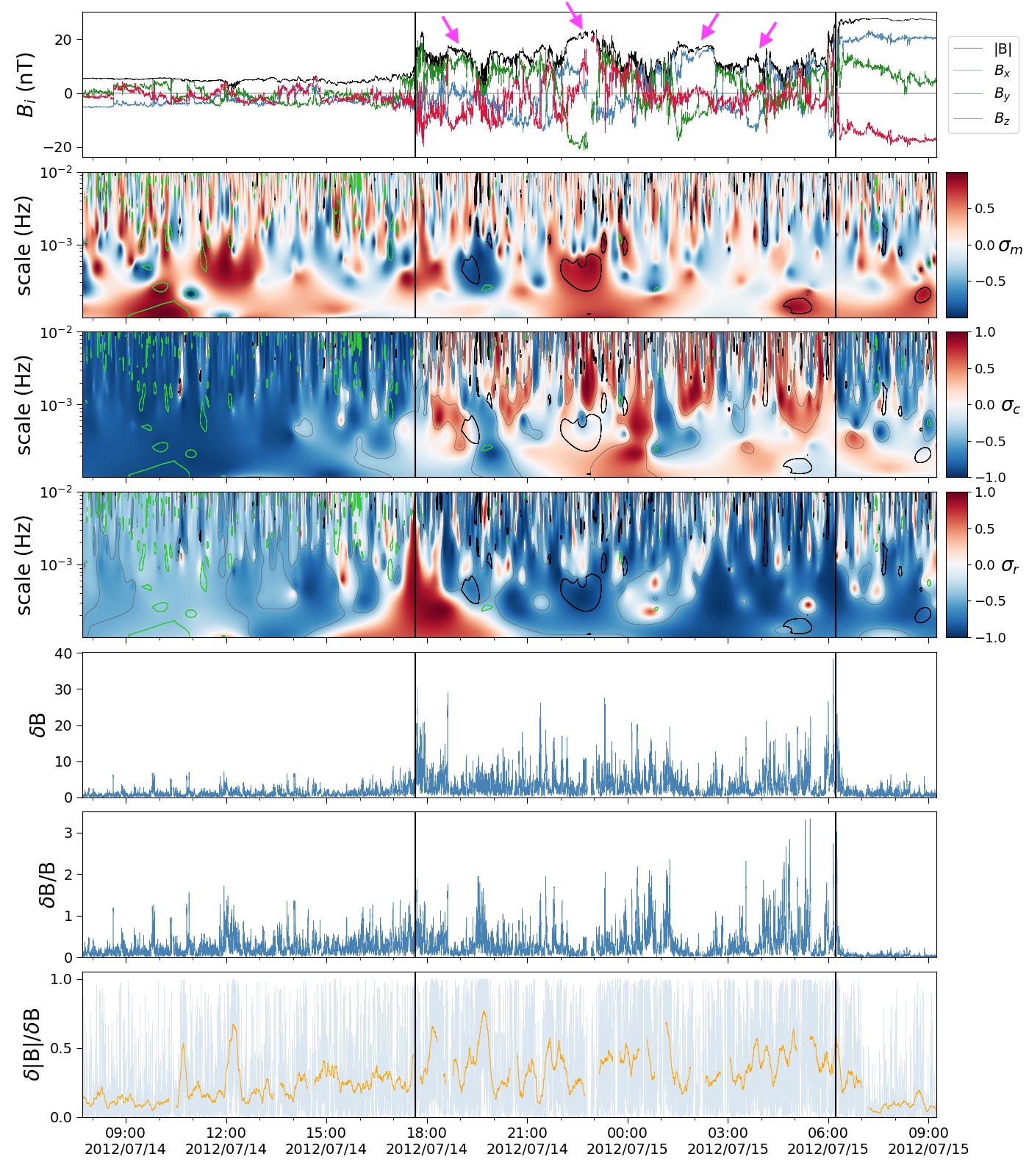}}
\caption{Sheath observed July 14--15, 2012. The panels show from top to bottom: (A) Magnetic field magnitude and components; (B) magnetic helicity; (C) cross helicity; (D) residual energy; (E) fluctuation amplitude; (F) normalized fluctuation amplitude; and (G) fluctuation compressibility and its 10-min average (orange curve). Gray contours show when the individual SFR criteria for $\sigma_m$, $\sigma_c$ and $\sigma_r$ are met, and black contours when they are all fulfilled. Green contours show periods when the criteria for Alfv\'en waves are all met. The sheath is bounded by the two vertical lines. Arrows point to weakly fluctuating magnetic field periods.}
\label{fig:example1}
\end{figure}

\begin{figure}   
\centerline{\includegraphics[width=0.99\textwidth,clip=]{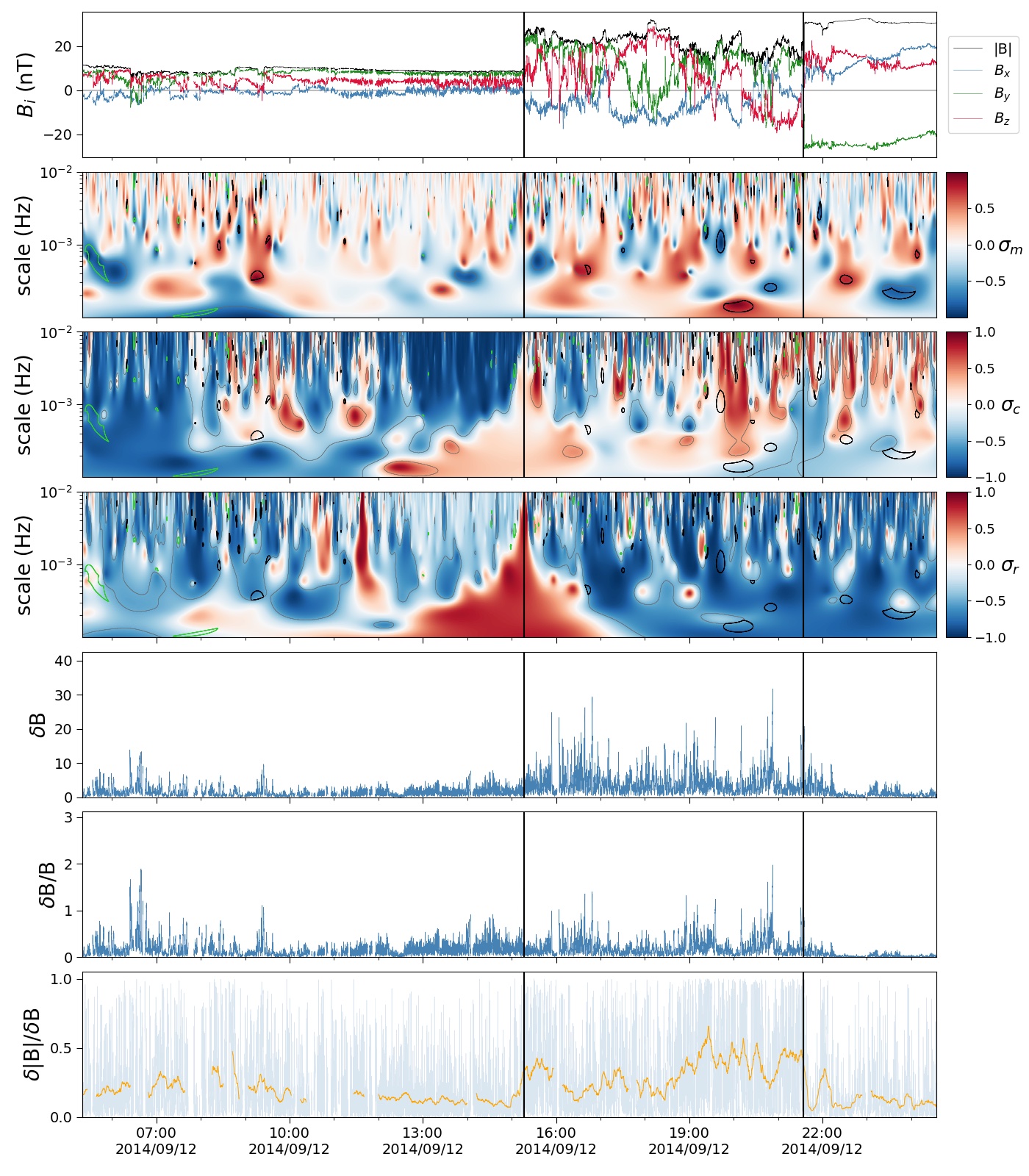}}
\caption{Sheath observed September 12, 2014, with panels in the same format as Figure~\ref{fig:example1}.}
\label{fig:example2}
\end{figure}

Two example events are shown in Figures~\ref{fig:example1} and \ref{fig:example2}. The panels give from the top to bottom the magnetic field magnitude and components in GSE coordinates, wavelet spectrograms of normalized magnetic helicity ($\sigma_m$), cross-helicity ($\sigma_c$) and residual energy ($\sigma_r$), fluctuation amplitude ($\delta B$), normalized fluctuation amplitude ($\delta B/B$) and compressibility of fluctuations ($\delta |B|/\delta B$). Gray contours show periods when the individual criteria of $\sigma_m$, $\sigma_c$ and $\sigma_r$ set for SFR identification (Section~\ref{subsec:methods}) were individually fulfilled, while black contours indicate when they were all fulfilled. Green contours similarly indicate Alfv\'{e}n wave occurrences.  A 1-min time lag was used to calculate $\delta B$, $\delta B/B$ and $\delta |B|/\delta B$. Fluctuations at this 1-min scale fall within the MHD range, but are at a smaller scale than the smallest time period of the SFRs studied here (1.67 min). 

The first example event occurred on July 14--15, 2012. The shock was observed at the Wind spacecraft on July 14, 17:39~UT and the ICME leading edge on July 15, 06:14~UT. The shock is marked by an abrupt increase of $|B|$, while the ICME leading edge occurred at the onset of a prolonged period of relatively smooth magnetic field. Note that only the first few hours of the ICME are shown here. The sheath in Figure \ref{fig:example1} features alternating periods of highly and weakly fluctuating magnetic field, the weak periods (see arrows in Figure \ref{fig:example1}) displaying lower $\delta B$ and $\delta B/B$. The first two weakly fluctuating periods are identified as SFRs at scales approximately $5 \times 10^{-4}$~Hz (30-min period). The two latter coherent structures are not identified as SFRs because these period do not fulfil the $\sigma_m$ criterion. There is also a low-frequency, $\sim$ 2.6-hr SFR just before the ICME leading edge at around 05:00~UT. From visual inspection, it does not appear to have a smooth flux rope rotation, and also exhibits high $\delta B$ and $\delta B/B$. Furthermore, the figure shows that SFRs, particularly at higher frequencies, are observed outside the intervals that appear smooth to the eye for the time-scales shown. They are observed throughout the sheath, but tend to occur in `trains' of several SFRs, i.e., there are periods with several high-frequency SFRs occurring with only a short time in between followed by a period devoid of SFRs. Visual inspection also indicates that SFRs appear more frequently in the sheath than in the upstream wind. For Alfv\'{e}n waves the opposite trend is seen, with Alfv\'{e}n waves more common in the upstream wind than in the sheath. Alfv\'{e}n waves are also more frequently occurring at high frequencies (small timescales). The three bottom panels of Figure \ref{fig:example1} indicate that fluctuation amplitudes and compressibility vary quite substantially through the sheath.  

The second example event occurred on September 12, 2014. The shock was observed at Wind at 15:17~UT and the ICME leading edge at 21:34~UT. Similarly to our first example, the shock is marked by a sharp increase in $|B|$ and the ICME leading edge by the transition to the large-scale coherent field of the ICME. Figure~\ref{fig:example2} exhibits periods of fluctuating and smoother magnetic fields in the sheath, like the sheath in Figure~\ref{fig:example1}. Only one lower frequency flux rope is identified in this sheath, occurring towards the latter part of the sheath at around $10^{-4}$~Hz ($\sim$ 2.5 hrs). The trailing part of the sheath had several SFRs in the medium frequency range at $\sim 5 \times 10^{-4}$~Hz ($\sim 30$ min), while they were less frequent in the leading part of the sheath. The occurrence of high-frequency SFRs was also greater in the trailing part than in the leading part of the sheath. Again, the upstream wind showed fewer SFRs, except in the period around September 12, 08:00-10:00~UT. In particular, the period just upstream of the shock completely lacked SFRs at all frequencies investigated. For our example event 2, there were no lower frequency Alfv\'{e}n waves in the sheath. However, some Alfv\'{e}n waves were observed at the highest frequencies throughout the sheath. Compared to the first example event, the upstream wind was less Alfv\'{e}nic at the frequencies we have analysed. The period before the shock was also lacking Alfv\'{e}n waves, but they occurred at the beginning of the interval shown. Again, the three bottom panels of Figure \ref{fig:example2} exhibit large variations throughout the sheath without obvious trends, except that the trailing part of the sheath where the majority of mid- and low-frequency SFRs in this case occurred had higher compressibility than the leading part. 

\begin{table}
    \centering
    \begin{tabular}{|c|c|c|c|c|c|c|} 
     \hline
     \multirow{2}{*}{} & \multicolumn{3}{c|}{Upstream wind} & \multicolumn{3}{c|}{Sheath}\\\cline{2-7}
      & high & mid & low & high & mid & low \\
     \hline
     \hline
     $\sigma_{m}$ & 23 & 27 & 33 & 24 & 32 & 40 \\ 
     \hline
     $\sigma_{c}$ & 51 & 52 & 56 & 69 & 73 & 81 \\
     \hline
     $\sigma_{r}$ & 66 & 76 & 75 & 76 & 82 & 80 \\
     \hline
     all & 4.6 & 6.6 & 8.3 & 8.2 & 13 & 16 \\
     \hline
    \end{tabular}
    \caption{Average occurrence percentages when individual identification criteria (see Section \ref{subsec:methods}) for SFRs are met (first three rows) and when they are simultaneously met (last row) for the upstream solar wind and sheath, and as a function of SFR frequency range (`high', `mid' and `low'). Values are rounded to two s.f.}
    \label{table:occurrence}
\end{table}

\subsection{Statistical results}
\label{sec:statistical}

Table~\ref{table:occurrence} shows the average percentage of time across all 55 events analyzed when the criteria for identifying SFRs was met, both in the upstream wind and sheath, and as a function of frequency range. These occurrence percentages are calculated for individual conditions as well as for times when all three conditions are simultaneously met. At a given time, a positive identification was made if the conditions were met at any frequency within the range under consideration. An upstream wind interval ahead of each sheath with duration 9~hr 57~min, equal to the mean  duration of the 55 sheaths, was considered. From the table, the same trends can be seen as for the two example events discussed in Section~\ref{sec:examples}, i.e., occurrence rates of SFRs are higher in the sheath than in the upstream for all three frequency ranges. It can also be seen that, of the three criteria, the $\sigma_m$ criterion is the least commonly met (and hence most stringent for SFR identification).

\begin{figure}   
\centerline{\includegraphics[width=0.8\textwidth,clip=]{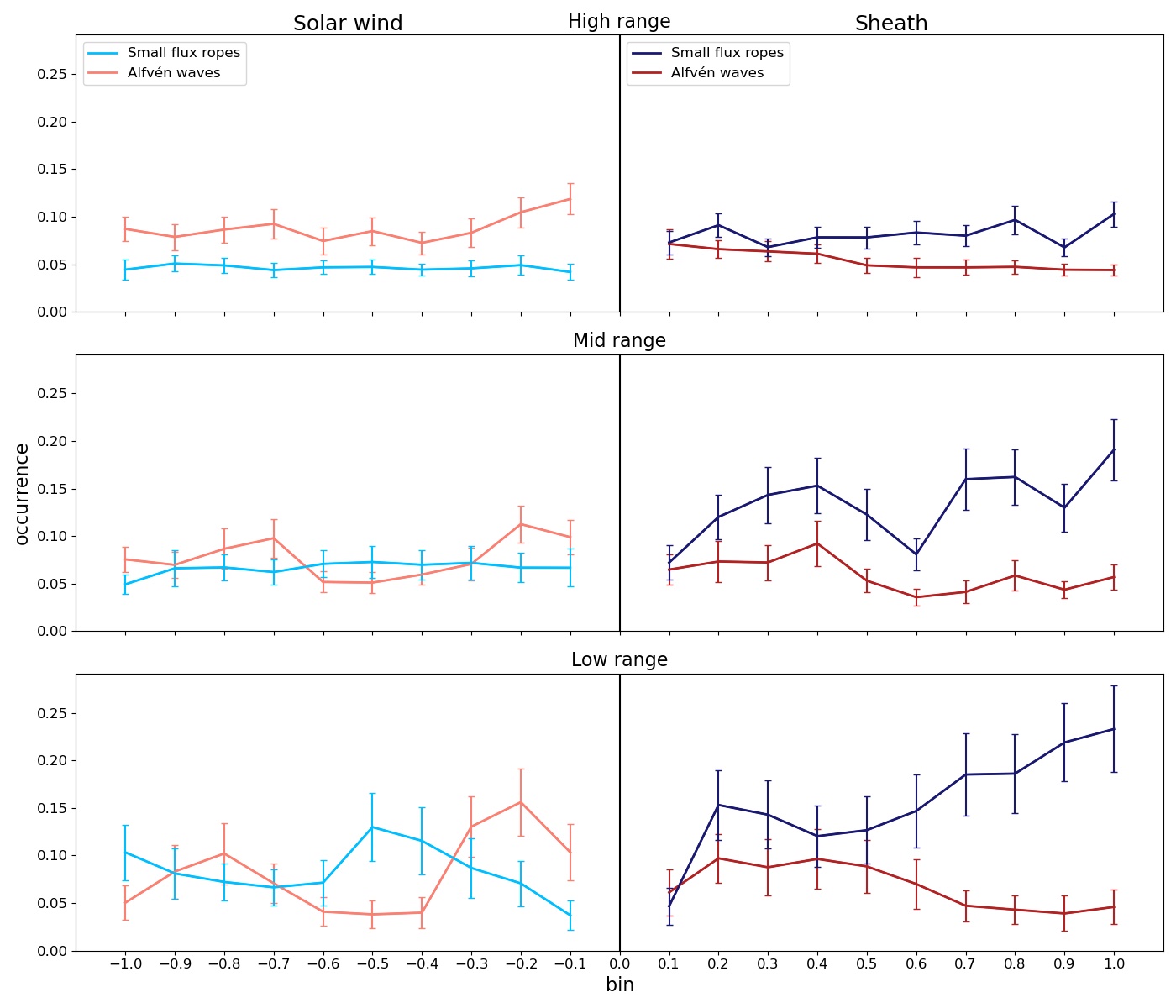}}
\caption{Superposed epoch profiles of SFR  and Alfv\'{e}n wave occurrence ratios in sheaths and the upstream wind for three SFR frequency ranges. On the x axis $-1$ corresponds to the start of the upstream wind period, 0 to the shock, and 1 to the leading edge of the ICME.}
\label{fig:occurrence}
\end{figure}

Figure~\ref{fig:occurrence} shows superposed epoch analysis of the SFR occurrence ratio, defined as the fraction of time SFRs are present, in the upstream wind and in the sheath. The upstream solar wind and the sheath intervals have each been divided into 10 bins in which the occurrence ratios have been calculated. As for the values in Table~\ref{table:occurrence}, a solar wind interval with duration 9~hr 57~min (equal to the mean duration of the sheaths) ahead of each sheath was analyzed. The figure shows averages of all events in each bin, with the standard deviation of the mean as error bars. The results are shown separately for the three frequency ranges. It can be seen that for all frequency ranges, SFRs appear to be more frequent in the sheath than in the upstream. Note that the average ratios across these profiles are consistent with the percentages in the bottom row of Table~\ref{table:occurrence}, as expected. The upstream wind occurrence profiles are fairly flat in the high- and mid-frequency ranges, while the low-frequency profile shows a bump and dip ahead of the shock. In the sheath, the occurrence rate of SFRs increases immediately behind the shock for the high-frequency range and then stays approximately flat. For the low-frequency range, the occurrence rate is approximately flat or rising towards the leading edge of the ICME, while the mid-frequency range is more variable. These profiles suggest that the smallest SFRs are evenly distributed in the sheath, while larger-scale SFRs tend to peak close the ICME leading edge. 

The occurrence rates of Alfv\'{e}n waves are shown in a similar manner in Figure~\ref{fig:occurrence}. In the high-frequency range in the upstream wind, the occurrence ratios of Alfv\'{e}n waves are considerably higher than the SFR ratios. The occurrence ratio peaks close to the shock. In the sheath, Alfv\'{e}n waves occur less commonly than SFRs, with their occurrence declining from the shock towards the ICME leading edge. For mid and low frequencies in the upstream wind, the occurrence ratios of Alfv\'{e}n waves are approximately similar to those of SFRs, while in the sheath they are clearly less frequent. At low frequencies, the occurrence ratios of Alfv\'{e}n waves in the upstream wind show a bimodal behaviour with the first peak at bins 3--4 and the second peak close to the shock. In the sheath at mid and low frequencies, the occurrence ratios also decline from the shock to the ICME leading edge. 

\begin{figure}   
\centerline{\includegraphics[width=0.8\textwidth,clip=]{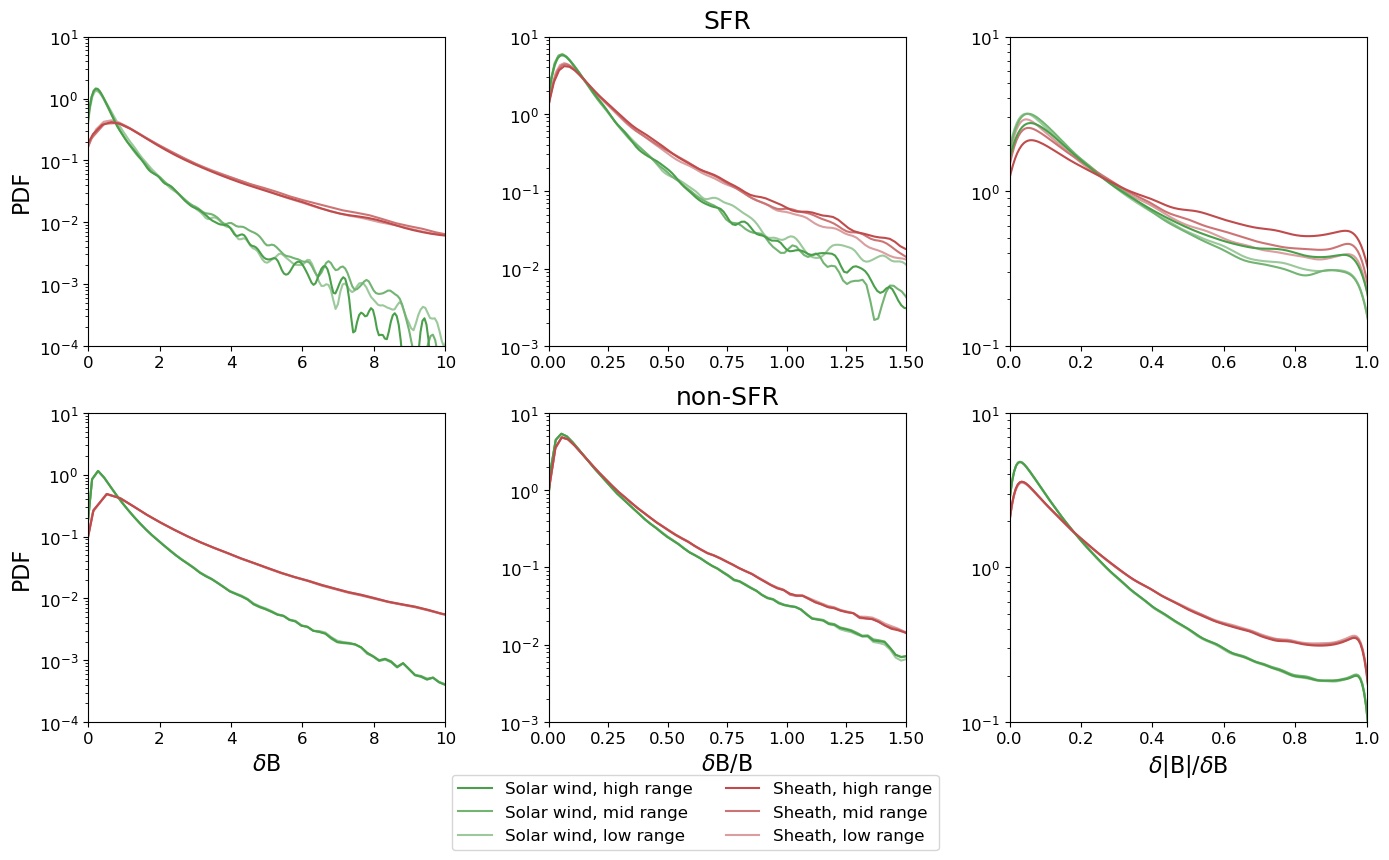}}
\caption{PDFs of magnetic field fluctuation amplitudes, normalized fluctuation amplitudes and compressibility at 1-min timescale (see text for details).}
\label{fig:PDF}
\end{figure}

Probability distribution functions (PDFs) of $\delta B$, $\delta B/B$ and $\delta |B|/\delta B$ at a timescale of 1~min are shown in Figure~\ref{fig:PDF} for all 55 sheaths (red curves) and upstream wind intervals (green curves). Separate distributions are shown for the times when SFRs were present (upper panels) and not present (lower panels) in the high, mid and low frequency ranges. The PDFs show no variation between different frequency ranges when SFRs were not present. For the periods when SFRs were present, there are some differences in the PDF tails. These differences are likely due to statistical noise in the case of $\delta B$ and $\delta B/B$. The differences in the $\delta |B|/\delta B$ PDFs during the SFR periods (top right panel), however, appear somewhat more significant; they suggest that fluctuations at a 1-min timescale are less compressible in larger SFRs (i.e. lower SFR frequency range). This trend of reducing $\delta |B|/\delta B$ with increasing flux rope size possibly continues up to the scale of large ICME flux ropes, which are known to contain fluctuations with low compressibility \citep[e.g.,][]{Kilpua2021energetic}.

The biggest differences are seen between PDFs in the sheath and in the upstream wind. Both SFR and non-SFR periods in the sheath clearly have higher fluctuation amplitudes. The normalized fluctuation amplitudes for SFR periods in the sheath peak approximately at the same value, but the sheath PDFs have fatter tails. The same is true for compressibility. Interestingly, there is not much difference between the SFR and non-SFR periods in the sheath: the PDFs are very similar for the fluctuation amplitudes, while the SFR periods are more compressible.

\section{Discussion and Summary} \label{sec:discussion}

We have performed a statistical analysis of small-scale flux ropes (SFRs) identified in 55 ICME sheaths observed by the Wind spacecraft in near-Earth space. The SFRs were identified using criteria set for the normalized magnetic helicity ($\sigma_m$), residual energy ($\sigma_r$) and cross-helicity ($\sigma_c$), following the approach outlined in \cite{Zhao2020} and \cite{zhao2021} for the first orbital encounters of Parker Solar Probe. In this approach, SFRs are identified as structures with low $\boldsymbol{v}-\boldsymbol{B}$ correlation (equivalent to low $|\sigma_c|$) at the scale of the SFR size, significantly greater energy in magnetic fluctuations than velocity fluctuations (strongly negative $\sigma_r$) and  helical magnetic fields (high $|\sigma_m|$).  We also investigated the occurrence of Alfv\'{e}n waves using the same set of parameters with different threshold criteria. Alfv\'{e}n waves were identified from time periods with high $\boldsymbol{v}-\boldsymbol{B}$ correlation (high $|\sigma_c|$), a balance of energy in magnetic and velocity fluctuations ($\sigma_r\sim 0$) and the same constraint on $\sigma_m$ as for SFR identification. We investigated the occurrence of SFRs and Alfv\'{e}n waves in three frequency ranges between $10^{-4}$ Hz ($\sim 167$ min) Hz and $10^{-2}$ Hz ($\sim 1.7$ min).

Using the above definitions, it was found that SFRs are frequently embedded within the sheath regions ahead of ICMEs at all frequencies (SFR scales) investigated. Conditions set for $\sigma_c$ and $\sigma_r$ to identify SFRs were individually fulfilled in the sheaths for the majority of the time ($\sim 70-80$\%), while the $\sigma_m$ criteria was met considerably less often ($\sim 20-40$\% of the time).

An interesting open question is whether SFRs are primarily generated in sheaths or compressed from the preceding wind into sheaths. We found that SFRs were more common in sheaths than in the upstream wind for all three frequency ranges investigated. The occurrence percentages were between $\sim 8\% - 16\%$ in the sheaths compared to $\sim 5\% - 8\%$ in the upstream wind. This trend was also clearly evident in both of our example events. The larger occurrence rate of SFRs in sheaths relative to the upstream wind suggests that, to some extent at least, SFRs form actively within ICME sheaths. The possible physical mechanisms could, as discussed partly in Section~\ref{sec:intro}, be generation via MHD turbulence \citep[e.g.,][]{Hu2018} and/or magnetic reconnection \citep[e.g.,][]{Feng2009,Zhao2020}. The level of turbulence is known to be enhanced in sheaths compared to the surrounding wind, and sheaths are expected to embed current sheets that act as favourable sites for reconnection to happen \citep{Kilpua2021stat}. 

In the upstream wind, the occurrence rate of SFRs was approximately steady, particularly at the high and mid-range frequencies. In the sheath, they also had approximately steady occurrence for the highest frequency range (smallest scale-size), but the occurrence rate increased towards the ICME leading edge for the mid and low frequencies. The sheath near the ICME leading edge is a highly dynamic region \citep[e.g.,][]{alalahti2019} where one process that could result in suitable current sheets for the generation of SFRs is field line draping around the ejecta \citep[e.g.,][]{gosling1987} that is associated with organized magnetic field structures \citep[e.g., planar magnetic structures;][]{Palmerio2016}. The draping gets more effective the faster the ICME propagates ahead and stronger it expands.  Furthermore, our results suggest that the ICME leading shock does not have much of a role in the generation of SFRs at lower frequencies. More insight on this question is expected from measurements by Parker Solar Probe \citep{fox2016} and Solar Orbiter \citep{muller2013}, as they detect SFRs and their generation much closer to the Sun than the Earth's orbit \citep[e.g.,][]{Chen2020PSP}. 

Differences in the SFR occurrence rates and their trends across the sheath between the low and high frequency SFRs could signify inherent differences between the dynamic processes creating and destroying them. A possible explanation for the flat profile at high frequencies and more rising profiles at mid and low frequencies is that small, high frequency SFRs are created and destroyed evenly throughout the sheath by local, small-scale processes \citep[e.g. reconnection of current sheets bounding turbulent eddies,][]{Hu2018}, while large, low frequency flux ropes are mostly pre-existing from the solar wind \citep[e.g.,][]{borovsky2020}, are more robust to erosion, and pile up towards the trailing part of the sheath. 

Our current study combines shocks of different Mach numbers and drivers with various properties. In general, the level of fluctuations is known to enhance at the transition from the upstream to the sheath, and more so when the sheath and driving ejecta are fast and the leading shock is strong/fast \citep[e.g.,][]{kilpua2020}. Therefore, one could expect that the enhancement of SFRs in the sheath with respect to the ambient wind (or at least, the population created spontaneously by the turbulent cascade) depends on the properties of the driver. To test the effect of shock strength, we compared SFR occurrence in sheaths with high and low magnetosonic Mach numbers ($M_{ms} > 2$ and $M_{ms} < 2$, respectively), but did not find a significant dependence on this shock parameter. The effect of shock properties on SFR generation and occurrence would merit a dedicated future study.

The occurrence rates of Alfv\'{e}n waves were relatively similar between the upstream wind and the sheath for the mid- and low-range frequencies, while for the highest frequencies there were more Alfv\'{e}nic fluctuations in the preceding wind than in the sheath. The rates in both domains peaked in the vicinity of the shock, i.e. the trend within the sheath was the opposite to that found for SFRs. This suggest that Alfv\'{e}n waves and SFRs are created by different physical processes and that shock-related processes are more efficient in generating Alfv\'{e}n waves than those in action at the ICME leading edge.

The recent study by \cite{Farrugia2020} found a sheath filled with Alfv\'{e}nic fluctuations. While this can be the case, our results highlight that there is a considerable sheath population with a significant amount of  non-Alfv\'{e}nic fluctuations embedded. 

An interesting finding was that the PDFs of fluctuation amplitudes, normalized fluctuation amplitudes and compressibility calculated using a 1-min time lag were nearly identical for different frequencies. This indicates that small-scale properties are independent of the size of the SFR embedded in the sheath or in the upstream wind. For both SFR and non-SFR periods, the PDFs  showed considerably fatter tails in the sheath than in the solar wind, in agreement with previous studies investigating the effect of this transition on fluctuation properties \citep{Kilpua2021stat}.

We also considered whether systematic changes in the values of $\sigma_m$, $\sigma_c$, and $\sigma_r$ between the upstream wind and sheath (e.g., due to the shock, or sheath compression) had any effect on the flux rope identification i.e. whether flux ropes identified in the sheath would not have been identified with the same threshold criteria in the upstream wind, due to such systematic changes. To test this, we found the average downstream-to-upstream ratios of $|\sigma_m|$, $|\sigma_c|$ and $\sigma_r$, which were ~1.07, ~0.804 and ~1.02, respectively. We note that that the $|\sigma_c|$ and $\sigma_r$ ratios are consistent with the analysis performed by \citet{good2022}. These ratios were calculated using the entire upstream wind and sheath intervals. The wavelet spectrograms for each upstream interval were scaled by these ratios, and the flux rope identification code was re-run across all of the rescaled upstream intervals. The occurrence percentages in the high, mid and low frequency bands rose to 6.0 \%, 11 \% and  13 \% respectively, higher than the unscaled upstream occurrence percentages listed in Table \ref{table:occurrence}, but not as high as in the sheath. These values suggest that at least some of the apparent increase in flux rope occurrence from solar wind to sheath is due to systematic changes in the values of $\sigma_m$, $\sigma_c$, and $\sigma_r$, rather than just creation of new flux ropes in the sheath. However, the test we have described here is not perfect, since any creation of new flux ropes will have some effect on the average values of $\sigma_m$, $\sigma_c$, and $\sigma_r$ in the sheath that were used to calculate the upstream-to-downstream ratios, in addition to other sheath-related processes.

To summarize, this study showed that SFRs are frequently present in ICME-driven sheath regions across a wide range of frequencies. They tend more often to occur close to the leading edge of the driving ejecta. Their small-scale fluctuation properties do not seem to depend on the size of the flux rope, being similar to that in the other parts of the sheath. The larger occurrence rate of SFRs in sheaths relative to the upstream wind implies that they are actively generated  within sheaths. Observations by Parker Solar Probe and Solar Orbiter are expected to bring new insights into the formation and evolution of SFRs. In addition, the effect of threshold parameters on flux rope identification is an interesting subject for future investigation. Some SFRs might lose their flux rope identity in interplanetary space \citep{Kilpua2009}, and therefore lowering the threshold for the magnetic helicity could help to identify such events.

\section*{Author Contributions}
JR performed the data analysis, produced the figures and contributed to the writing of the paper. EK took the main responsibility for writing the introduction and discussion. All authors contributed to the planning of the analysis, physical interpretations, and to the writing of the paper.

\section*{Funding}
This work was supported by the SolMAG project ERC-COG 724391 funded by the European Research Council's Horizon 2020 Research and Innovation Programme, and by the Academy of Finland (AoF) project SMASH, grant 310445. The results presented here have been achieved under the framework of the Finnish Centre of Excellence in Research of Sustainable Space, AoF grant 312390, which we gratefully acknowledge. SWG is supported by AoF grants 338486 and 346612. MA-L acknowledges the Emil Aaltonen Foundation for financial support.

\section*{Acknowledgments}  
We thank the Wind/MFI and Wind/3DP instrument teams for the data used in this study.

\section*{Supplemental Data}

\section*{Data Availability Statement}
Wind data are publicly available at NASA's Coordinated Data Analysis Web (CDAWeb) database (\url{http://cdaweb.gsfc.nasa.gov/}). The ICME catalogue by \citet{Nieves_Chinchilla2018} is available at \url{https://wind.nasa.gov/ICMEindex.php} and HELCATS ICME catalogue at \url{https://www.helcats-fp7.eu/catalogues/wp4_icmecat.html}.

\bibliographystyle{frontiersinSCNS_ENG_HUMS} 
\bibliography{SmallFR_bib}





\end{document}